\documentclass[%
twocolumn,
superscriptaddress,
 aps,
prl,
]{revtex4-1}

\usepackage[utf8]{inputenc}
\usepackage[T1]{fontenc}
\usepackage[colorinlistoftodos, color=green!40, prependcaption]{todonotes}

\usepackage{CJK}

\usepackage{graphicx}
\DeclareGraphicsExtensions{.pdf,.png}
\usepackage{capt-of}
\usepackage[colorlinks=true, citecolor=blue, linkcolor=blue]{hyperref}
\usepackage[colorinlistoftodos]{todonotes}

\begin{document}


\title{Focussing frustration for self-limiting assembly of flexible, curved particles}

\author{Nabila Tanjeem}
\affiliation{Department of Chemical and Biological Engineering, University of Colorado, Boulder, Colorado 80303, USA}
\author{Douglas M. Hall}
\affiliation{Department of Polymer Science and Engineering, University of Massachusetts, Amherst, Massachusetts 01003, USA}
\author{Montana B. Minnis}
\affiliation{Department of Chemical and Biological Engineering, University of Colorado, Boulder, Colorado 80303, USA}
\author{Ryan C. Hayward}
\affiliation{Department of Chemical and Biological Engineering, University of Colorado, Boulder, Colorado 80303, USA}
\author{Gregory M. Grason}
\affiliation{Department of Polymer Science and Engineering, University of Massachusetts, Amherst, Massachusetts 01003, USA}

\begin{abstract} 
We show that geometric frustration in a broad class of deformable and naturally curved, shell-like colloidal particles gives rise to self-limiting assembly of finite-sized stacks that far exceed particle dimensions. When inter-particle adhesions favor conformal stacking, particle shape requires {\it curvature focussing} in the stack, leading to a super-extensive accumulation of bending costs that ultimately limit the ground-state stack size to a finite value. Using a combination of continuum theory and particle-based simulation, we demonstrate that the self-limiting stack size is controlled by the ratio of the intra-particle bending costs to inter-particle adhesion energy, ultimately achieving assembly sizes that are tuned from a few, up to several tens of, particles.  We show that the range of  self-limiting assembly is delimited by the two structural modes of ``frustration escape'' which evade the thermodynamic costs of curvature focussing. Crucially, each of these modes can be suppressed through suitable choice of adhesive range and lateral patchiness of adhesion, providing feasible strategies to program finite assembly size via the interplay between shape-frustration, binding and deformability of colloidal building blocks.

\end{abstract}

\maketitle


Advanced methods to control the size, shape and interactions of synthetic particles \cite{li_assembly_2016,vogel_advances_2015,li_colloidal_2011,li_colloidal_2020,heuckel_2021} continue to drive remarkable progress in formation of hierarchical structures via colloidal assembly. Yet, the prevailing paradigms almost exclusively target nearly strain-free structures, whose equilibrium dimensions grow to uncontrolled sizes to minimize free energy. In contrast, several recent studies and models point to the possibility of exploiting the size-dependent costs of geometric frustration to control the equilibrium finite size and shape of assemblies \cite{grason_perspective_2016, meng_elastic_2014,hall_morphology_2016, lenz_geometrical_2017,berengut_self-limiting_2020,meiri_cumulative_2021,tyukodi_thermodynamic_2021,serafin_frustrated_2021,zhang_shape_2019,yang_self-limited_2010}. Finite assemblies may enable useful mimics of size-limited natural structures, such as viral capsids \cite{baker_adding_1999,zlotnick_virus_2011,zandi_virus_2020}, bacterial microcompartments \cite{kerfeld_bacterial_2010,chowdhury_diverse_2014}, structurally colored protein superstructures \cite{prum_development_2009,dufresne_self-assembly_2009,saranathan_structure_2012}, and multi-filament bundles \cite{weisel_structure_2007,popp_supramolecular_2012}. 

Self-limitation in frustrated systems occurs when local misfits between the shapes of subunits 
incur elastic costs for assembly that accumulate superextensively with assembly size \cite{grason_perspective_2016}. When those costs balance cohesive interactions, they define a thermodynamically selected finite size that can, in principle, substantially exceed the sizes of subunits or their interaction range. Self-limitation implies a minimum in the free energy per subunit at finite aggregation number, which implies  {\it pseudo-critical aggregation transition} to a state dominated by finite aggregates at high enough subunit concentration~\cite{hagan_equilibrium_2021}.  To date, understanding of this basic paradigm derives almost exclusively from continuum elastic theories, where the elastic costs and magnitudes of frustration are phenomenological parameters, and intra-assembly stresses are modeled in simplified morphologies \cite{schneider_shapes_2005,ghafouri_helicoid_2005,armon_shape_2014,grason_chiral_2020,meng_elastic_2014, hall_morphology_2016,serafin_frustrated_2021}.  As such, these models fail to survey the low-energy, symmetry-breaking modes of ``frustration escape'' by which physical assemblies evade the costs of accumulating frustration.  The few ``discrete subunit'' models of frustrated assembly studied so far consider only minimal descriptions of elastic polygons with infinitely short-ranged binding interactions, and coarse-grained (i.e. vertex based) elasticity models \cite{lenz_geometrical_2017,meiri_cumulative_2021,tyukodi_thermodynamic_2021}.  As self-limiting assembly derives from a complex interplay between particle shape, interactions, and deformability, such models leave open key basic questions: what is the accessible range of self-limiting assembly for a given frustrated particle design? How are self-limitation, or frustration escape, controlled by physical properties of particles? 

In this letter, we demonstrate the design of frustrated colloidal particles that exhibit tunable self-limiting assembly sizes. The misfitting subunits are deformable, curved elastic shells (dubbed `curvamers') that stack face-to-face due to short-ranged attractions. Uniform spacing between conformally-contacting curvamers, however, yields gradients of local curvature along the stacks (Fig. 1b), a phenomenon well known in focal conic domains of liquid crystals and geometric optics \cite{sethna_spheric_1982,fournier_geometrical_1996,didonna_smectic_2003,nye_natural_2000}, but which here provides a mechanism to propagate frustration to especially large inter-particle scales (i.e. $\gg 1$). Notably, we show by a combination of analytical theory and coarse-grained particle simulations that careful design of the ratio of elastic costs of particle bending and inter-particle adhesion leads to controlled self-limitation up to at least several 10s of curvamers.  We further establish that frustration escape, leading to ground state structures of unlimited size, can occur through two distinct mechanisms of ``curvature defocussing'', but that self-limitation can be maintained through suitable control of interaction range and patchiness, design principles that should be readily achievable with a range of synthetic colloidal building blocks.

 \begin{figure}
    \centering
    \includegraphics[width=8.6 cm]{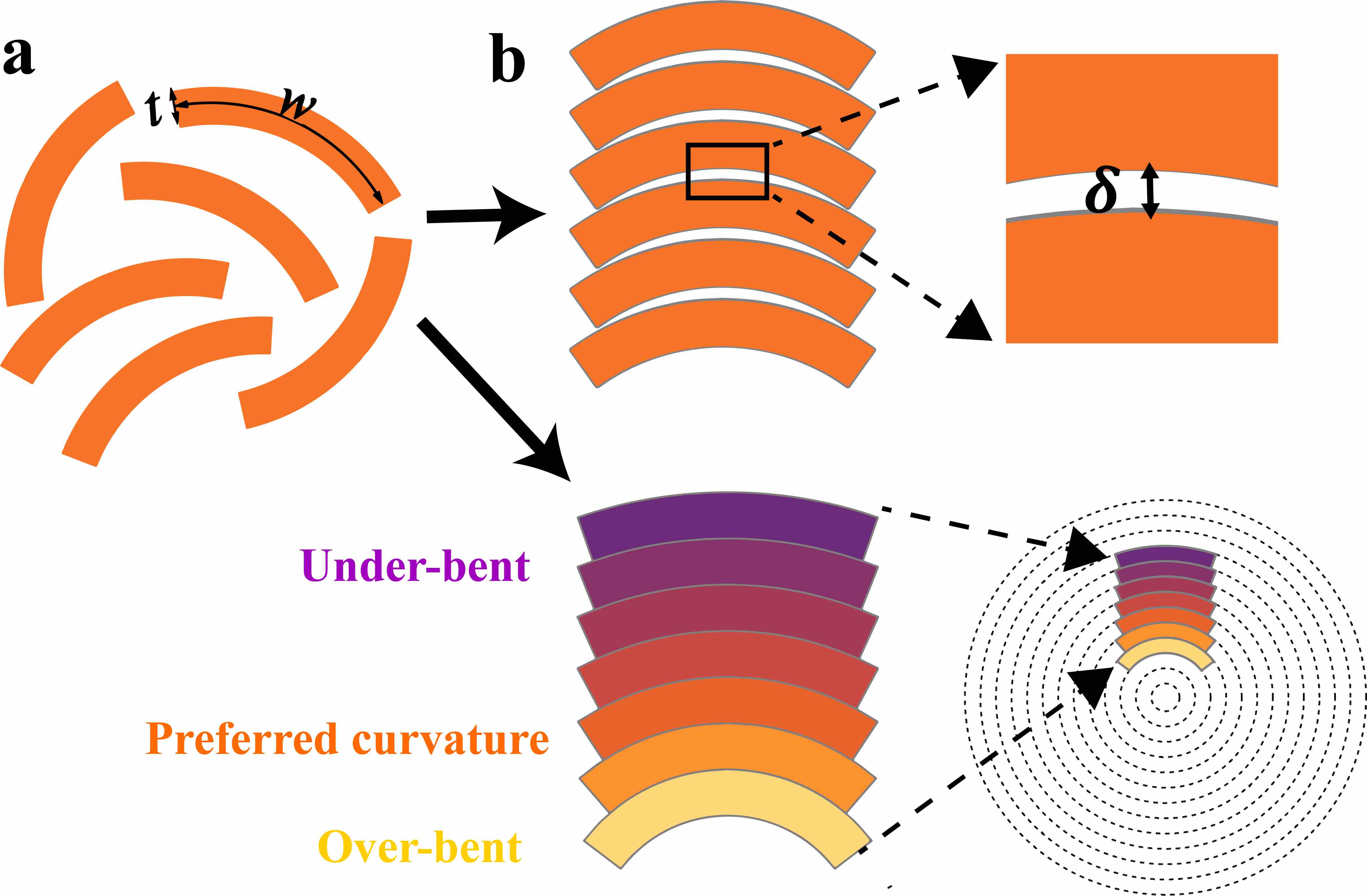}
    
    \captionof{figure}{\textbf{(a)} Curvamers with thickness $t$, width $w$ and preferred curvature radius $r_0$. \textbf{(b)} Top: Stacking of undeformed curvamers with a nominal separation distance $\delta$.  Bottom: Concentric stacking of flexible curvamers achieving conformal contact at the expense of intra-stack shape gradients.  }
    \label{fig:Curvamer_theory}
 \end{figure}
 

We begin by developing a simplified analytical theory of assembly of cylindrically curved shells, which we model in terms of their 2D cross-sections, ignoring distortions along their axial direction. As shown in Fig.~\ref{fig:Curvamer_theory}a, the geometry of each curvamer is defined by a preferred curvature $\kappa_0$ (i.e. $r_0 = \kappa_0^{-1}$ is the radius of curvature of the midline), thickness $t$, and width $w$. If curvamers retain their preferred shape, non-overlap between shells requires a gap between curvamer surfaces, which is maximal at their centers $\delta \approx t w^2 \kappa_0^2 / 8$. The condition of face-to-face assembly with a perfect, conformal contact requires the curvamers to bend and deviate from a fixed shape (Fig.~\ref{fig:Curvamer_theory}b).  To stack the ($n+1$)th curvamer on top of curvamer $n$, perfect contact requires a concentric stacking, or $r_{n+1}- r_{n} =t$, or
\begin{equation}
    \kappa_{n+1}=\frac{\kappa_n}{1+\kappa_n t}.
    \label{eq: focus}
\end{equation}
This is the condition of {\it curvature focussing} required by constant spacing between curved surfaces, which introduces shape gradients at the expense of elastic costs of over- and under-bending of particles. The energy of a stack of $N$ curvamers is the sum of the adhesive gain ($E_{\rm co}$) and elastic bending costs ($E_{\rm el}$),
\begin{equation}
\label{eq:total_energy}
E(N) =  E_{\rm co} (N)+E_{{\rm el}} (N)= - \gamma A (N-1)+ \frac{B A}{2} \sum_{n=1}^N \big( \kappa_n - \kappa_0\big)^2  ,
\end{equation}
 where $\gamma$ and $B$ are the adhesive energy (for perfect contact) per unit area and curvamer bending modulus, respectively, and $A$ the area of curvamer surfaces. To understand the mechanism of self-limitation, we consider the limit of stacks that are short compared to $r_0$, which can be analyzed in terms of the curvature at a layer height $z$ relative to a mid-layer of preferred shape, which according to eq. (\ref{eq: focus}) exhibit a linear variation in bending $\kappa(z) \simeq \kappa_0-\kappa_0^2 z$.  Averaging bending cost over the stack of size $N$, we expect $E_{{\rm el}} (N) \approx B A \kappa_0^4 t^2 N^3/24$.  Assemblies are dominated by the aggregates that minimize the free energy per subunit.  Taking $E(N) /(A N) \approx-\gamma + \gamma/N +  B \kappa_0^4 t^2 N^2/24 $, we thus expect a selected size $N_{\rm min} \approx (12 \gamma/B\kappa_0^2)^{1/3} (\kappa_0 t)^{-2/3} $.  A more complete analysis of the continuum model (see SI, sec. 1) shows that this power-law growth of the self-limiting stack persists up to a size $N_{\rm min} \lesssim r_0/t$ beyond which the selected sizes grow more rapidly as the mean curvature of all particles begins to flatten with stack growth.  Defining the dimensionless adhesion-to-bending ratio $S \equiv \gamma t / B \kappa_0$,
 and scaled stack size $H = N \kappa_0 t$, the self-limiting size of conformal stacks satisfies the equation of state (SI eq. S10)
 \begin{equation}
    S(H_*) = \frac{2-2\sqrt{H_*^2 +1}}{H_*}+ \sinh^{-1} H_*
    \label{eq: Hstar}
\end{equation}
According to this relationship, the power-law scaling for small stack size, $H_*(S\ll1)\sim S^{1/3}$, gives way to exponential growth $H_* (S\gg1) \sim e^{S}$ for large $S$.
 
\begin{figure}
  
    \centering
    \includegraphics[width=8.6 cm]{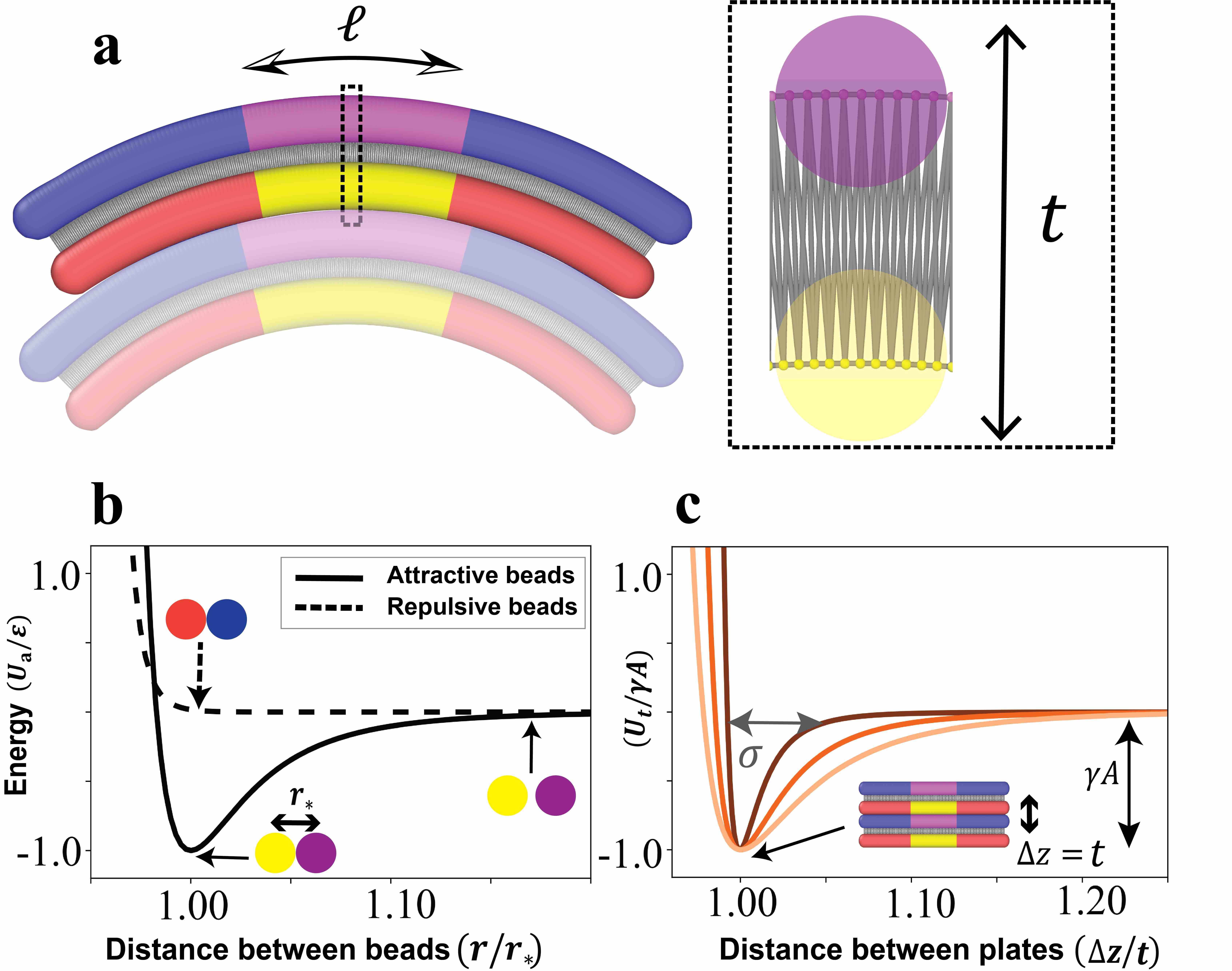}
    \caption{\textbf{(a)} Schematic of the bead-spring construction of the discrete curvature model.  Interaction sites are labeled as colored beads connected by a truss network of harmonic springs, depicted as grey bars.   \textbf{(b)} Pairwise potential between the red-blue bead pairs and the yellow-magenta bead pairs.  The attractive well depth is almost entirely due to the attractive yellow-magenta LJ patches, with red-blue flanks generating pure repulsion (via a WCA-like potential) beyond the attractive minimum, and range of attraction chosen to be much smaller than thickness ($\sigma\leq 0.18t$). \textbf{(c)} Total adhesion energy between two (flat) particles as function of scaled center-to-center spacing for different range of attractive interaction $\sigma = 0.06t, 0.12t,0.18t$.}
    \label{fig:lammps_model}
 \end{figure}

The continuum model, which assumes perfect curvamer alignment and conformal contact, predicts that the self-limiting stack size grows arbitrarily large with increasing $S$.  To test the limits of self-limitation, specifically the ability of more complex relaxation modes of curvamer stacks to circumvent frustration, we turn to numerical simulations (implemented in LAMMPS~\cite{LAMMPS}) of a 2D coarse-grained curvamer model. We implement elastic shell mechanics via a bead-spring truss network (see Fig.~\ref{fig:lammps_model}a and sec. 2 of SI for more details), with rest lengths chosen to set the bottom and the top layer radii of curvature to $r_{\rm in}=r_0-\frac{t_0}{2}$ and $r_{\rm out}=r_0+\frac{t_0}{2}$, respectively. The isosceles trapezoidal unit cell of this bilayer structure (Fig.~\ref{fig:lammps_model}a) comprises horizontal, vertical and diagonal springs with stiffnesses set to achieve the bending mechanics of an elastic shell of thickness $t_0$ and Poisson ratio 0.3 (Fig. S3).

To model attractive interactions between curvamers, we parameterize pair-wise interactions between vertices on the inner and outer faces of two types:  finite range attraction for an inner patch of width $\ell\leq w$ (magenta and yellow beads in Fig. 2a) and pure repulsion in the outer flanks of the particle (blue and red beads in Fig. 2a).  
The pairwise attraction between beads in the inner patch is given by a shifted Lennard-Jones (LJ) potential 
\begin{equation}
    U_a(r) = 4 \epsilon \left [ \left (\frac{\sigma}{r-\Delta}\right)^{12} -\left (\frac{\sigma}{r-\Delta}\right)^{6} \right],
    \label{eq:LJ}
\end{equation}
  where $\sigma$ is the range of the attractive well and $\Delta$ is a shift  parameter that controls the equilibrium separation $r_*=0.71t_0$ between attractive sites independent of $\sigma$.  The LJ potential (with beads placed at a high linear density, $\lambda = 16.1 t_0^{-1}$) is designed to model favorable uniform and frictionless contact with center-to-center spacing between bound curvamers $t=t_0+r_*$.  Outside of this attractive zone, repulsive interactions (with beads at the same density $\lambda$) are modeled by a Weeks-Chandler-Anderson (WCA) like potential whose smooth cut-off is matched to the minimum of $U_a(r)$ at $r_*$, such that at perfect conformal contact, repulsive sites do not contribute to the net interaction energy between bound curvamers (the repulsion strength is set to $10^{-3} \epsilon$, with $\epsilon$ the attractive strength in equation \ref{eq:LJ}).  We define $- \gamma A$ as the total attractive potential between two conformally contacting curvamers, and compute it as the depth of attractive interactions for two planar particles (i.e. flattened shells) shown in Fig.~\ref{fig:lammps_model}c.  We expect a dependence of $ \gamma \propto \epsilon \ell  \sqrt{r_* \sigma} \lambda^2 $, as each site interacts with a number $\propto \sqrt{r_* \sigma} \lambda$ on the opposing particle face, and the surface-surface interaction range is close to $\sigma$ (Fig.~\ref{fig:lammps_model}c).  To map coarse-grained curvamer parameters to the dimensionless adhesion $S$, we assume $\gamma$ to be independent of particle curvature, and further measure the shell stiffness $B$ by computing curvamer energy for variable circular curvatures of the midline (see Figs. S2 and S3).
  
  \begin{figure}
    \centering
    \includegraphics[width=8.6 cm]{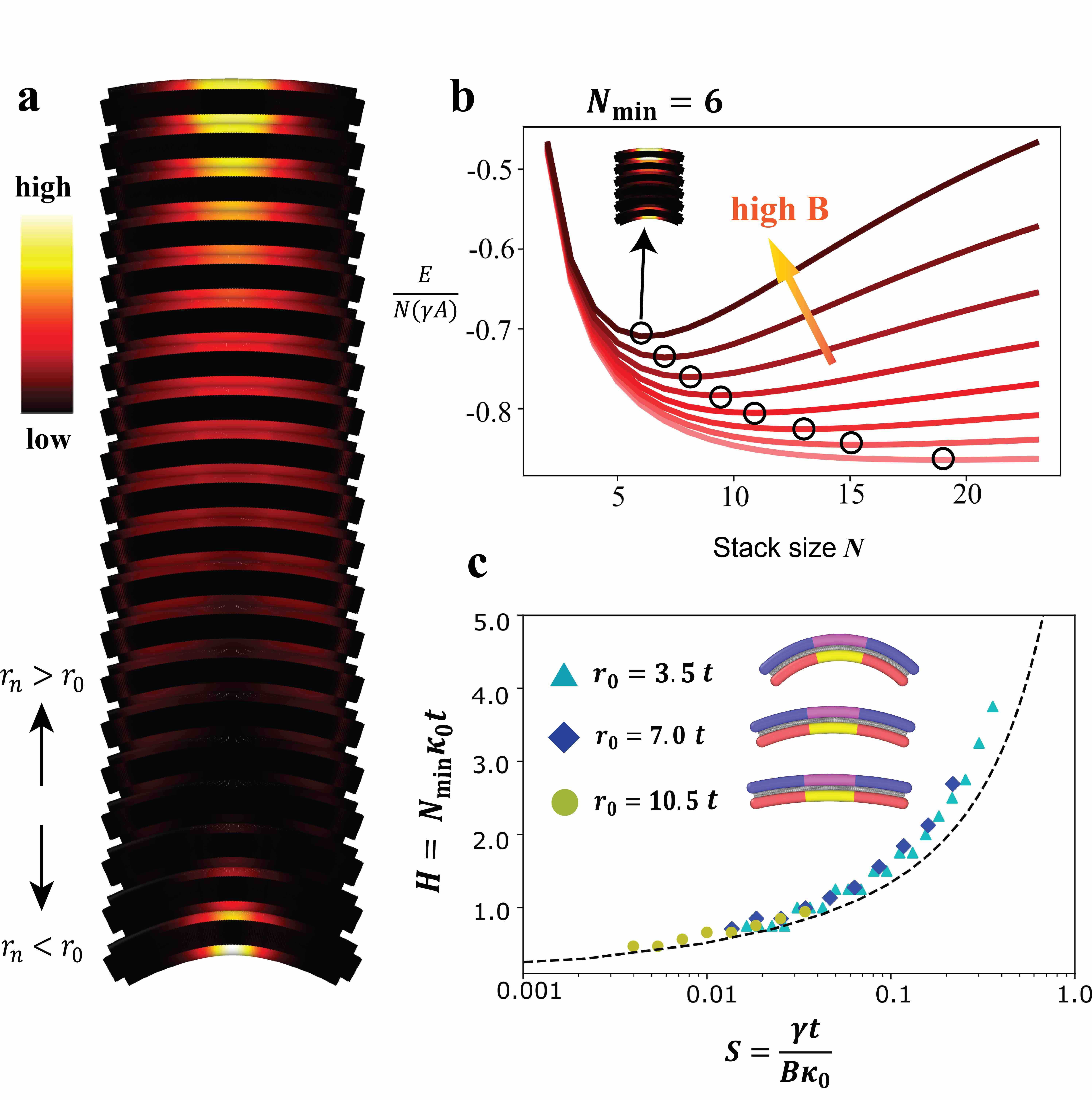}
    \caption{\textbf{(a)} Bond energy map for a stack with $N=20$ at the ground state, with bright colors representing higher energy bonds. Over/underbending deformations are most prominent at the ends of the stack and result in differential stretching/compression of horizontal bonds on the top and bottom of particles. \textbf{(b)} Plots of normalized energy per curvamer as a function of stack size, where darker colors represent curvamer assemblies with higher bending modulus. Range of energy ratio ${\gamma}/({B{\kappa_0}^2})$ and $S$ was varied between $0.09-1.44$ and $0.01-0.22$, respectively (see design 2 on table S6) \textbf{(c)} Relationship between the dimensionless parameters $S$ and $H$  for three different curvamer geometries with $\kappa_0 t = 0.29,0.14,0.09$. The dashed line is calculated from the continuum model.  }
    \label{fig:S-H}
 \end{figure}

To assess the assembly energy landscape via a particle-based model of curvamers, we perform energy minimization for stacks of $N$ curvamers, analyzing first the case of $\ell = w/3$,  $\sigma = 0.06t$ and  $r_0 = 7.0t$. Beginning from a (curvature focussing) configuration of concentrically stacked particles, energy is relaxed via a conjugate gradient algorithm in LAMMPS.  The thermal stability of these ground states was studied via simulated annealing as described in the SI, sec.5. Fig.~\ref{fig:S-H}a shows the energy density in the horizontal springs due to variable curvature through the stack thickness (Fig.S8) for the ground state of an $N=20$ curvamer stack. The energy per curvamer plotted in Fig.~\ref{fig:S-H}b as a function of the stack size for a sequence of increasing bending stiffness (corresponding to $S=0.01-0.22$) shows a global minimum at $N_{\rm min}$, indicative of self-limiting stack assembly~\footnote{$N_{\rm min}$ corresponds to the dominant aggregate size in the canonical ensemble, well above the critical aggregation concentration~\cite{hagan_equilibrium_2021}.}. Notably, the energy minimum shifts to smaller $N_{\rm min}$ with increased bending stiffness (Fig.~\ref{fig:S-H}b). 

These results confirm that the self-limitation derives from the accumulated bending strain generated via curvature-focussing stacking geometry, and further that the equilibrium stack size {\it decreases} with that elastic cost.  In Fig.~\ref{fig:S-H}c we compare the optimal stack sizes from the discrete curvamer model to the continuum results in eq. (\ref{eq: Hstar}) for uniform conformal contact. Considering more than two orders of magnitude in dimensionless adhesion to bending stiffness ratios $S$, we find that results from three different particle curvatures collapse onto a single curve whose monotonic increase with $S$ shows good agreement with the conformal contact model, notwithstanding the fact that equilibrium shapes of discrete curvamers (Fig.~\ref{fig:S-H}a) deviate considerably from the idealized circular shapes assumed in the model.


   \begin{figure}
    \centering
    \includegraphics[width=8.6 cm]{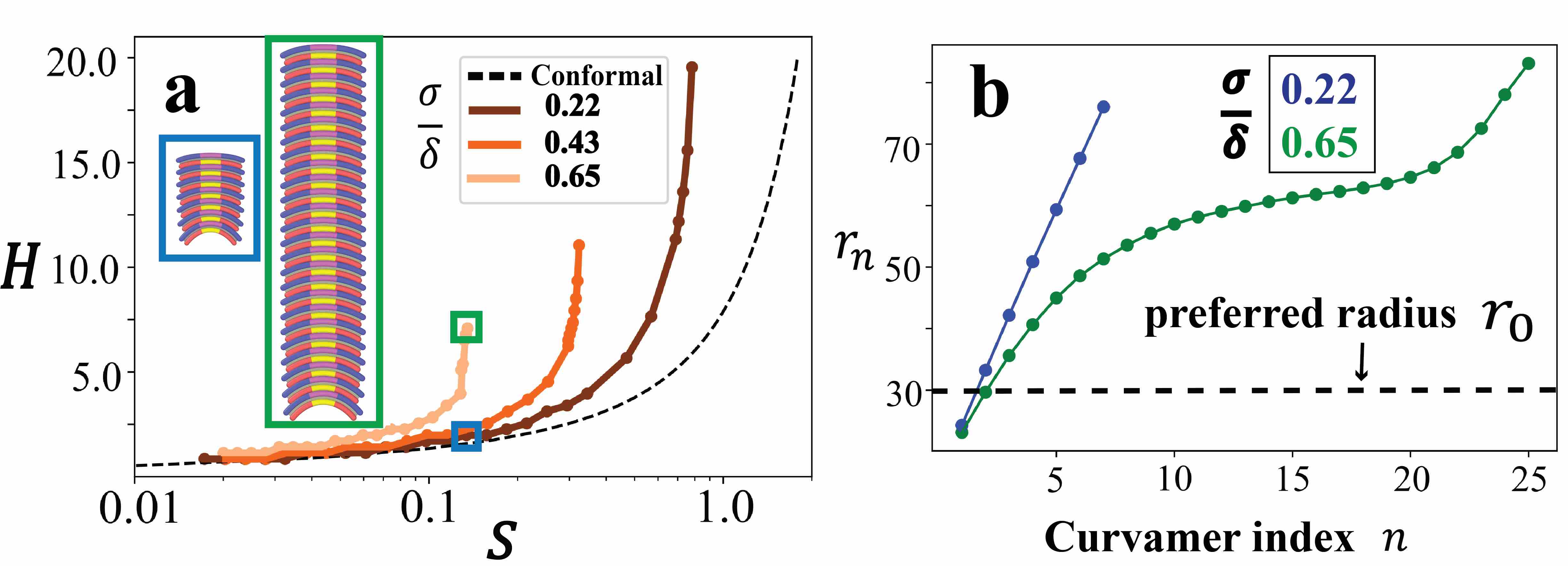}
    \caption{\textbf{(a)} Self-limiting stack size vs. dimensionless adhesion for three different ratios of attraction range ($\sigma$) to nominal gap size ($\delta$). The two stacks on the inset represent the self-limiting stacks for $\frac{\sigma}{\delta} = 0.22$ and $\frac{\sigma}{\delta} = 0.65$ at the same dimensionless adhesion indicated by the small blue and green squares, respectively.\textbf{(b)} The radii of curvature ($r_n$) of each curvamer in the two stacks plotted against their position index $n$. The dashed line shows the preferred radius $r_0$.
    }
    \label{fig:escape1}
 \end{figure}
 
In the cases summarized in Fig. ~\ref{fig:S-H}, self-limitation relies on the propagation of curvature focussing from one end of the stack to the other, leading to self-limiting sizes that far exceed the single building block size (i.e. $N_{\rm min} \gg 1$) and can vary by at least an order of magnitude with the ratio of inter-particle binding to intra-particle stiffness.  Next, we investigate two mechanisms through which assemblies can ``defocus'' curvature propagation and escape the cumulative costs of frustration.

  The first mode of escape is observed when adhesive interactions between curvamers are sufficiently long-ranged.  Intuitively, this can arise when $\sigma$ is much larger than the nominal gap size $\delta$ between undeformed curvamers, in which case the pair maintains strong adhesion without shape change, or its elastic cost. Fig.~\ref{fig:escape1} shows results for optimal stack sizes for curvamers of constant shape ($r_0 = 3.5t$), but varying ratio of adhesive range to nominal gap size, $\sigma/\delta$.  The equilibrium stack size generally exceeds the values predicted for perfect conformal contact, but also increases with the interaction range for a fixed dimensions adhesion $S$.  For example, for $S=0.14$, the optimal stack grows from $N_{\rm min}= 7$ for $\sigma/\delta = 0.22$ to $N_{\rm min}= 25$ for $\sigma/\delta = 0.65$ (Fig.~\ref{fig:escape1}a).  Further, in clear distinction to the conformal contact model, which predicts self-limiting stacks for all $S$, we observe an upper limit to the adhesion strength $S_{\rm max}$ above which no minimum in $E(N)/N$ can be identified, which decreases with increasing $\sigma$ (Fig.~\ref{fig:escape1}a and Fig. S10).  To explain these effects, we compare the shape profiles of optimal curvamer stacks with two values of interaction range, $\sigma/\delta = 0.22$ and $0.65$ in  Fig.~\ref{fig:escape1}b.  While shorter range interactions (blue curve) yield curvature radii that increase roughly linearly with $n$, corresponding to curvature-focussing, longer range interactions (green curve) provide a slower and non-linear increase of curvature radius along the stack.  This non-linear profile indicates the opening of a small gap between curvamer faces (see SI Fig. S9), allowing interior curvamers to maintain relaxed and roughly constant curvature shapes closer to the preferred shape.  Hence, the longer range interactions shift the effect of frustration to strain inter-particle bonds, weakening the effect of curvature frustration.  Thus, as $\sigma$ increases for a fixed $S$, the cumulative elastic costs of frustration are reduced, allowing optimal stacks to reach larger sizes.  Ultimately, the assembly ``escapes'' frustration when the accumulating elastic costs of curvature focussing overwhelm the cost of uniform shape, gap-strained stacking.
  
     \begin{figure}
    \centering
    \includegraphics[width=8.6 cm]{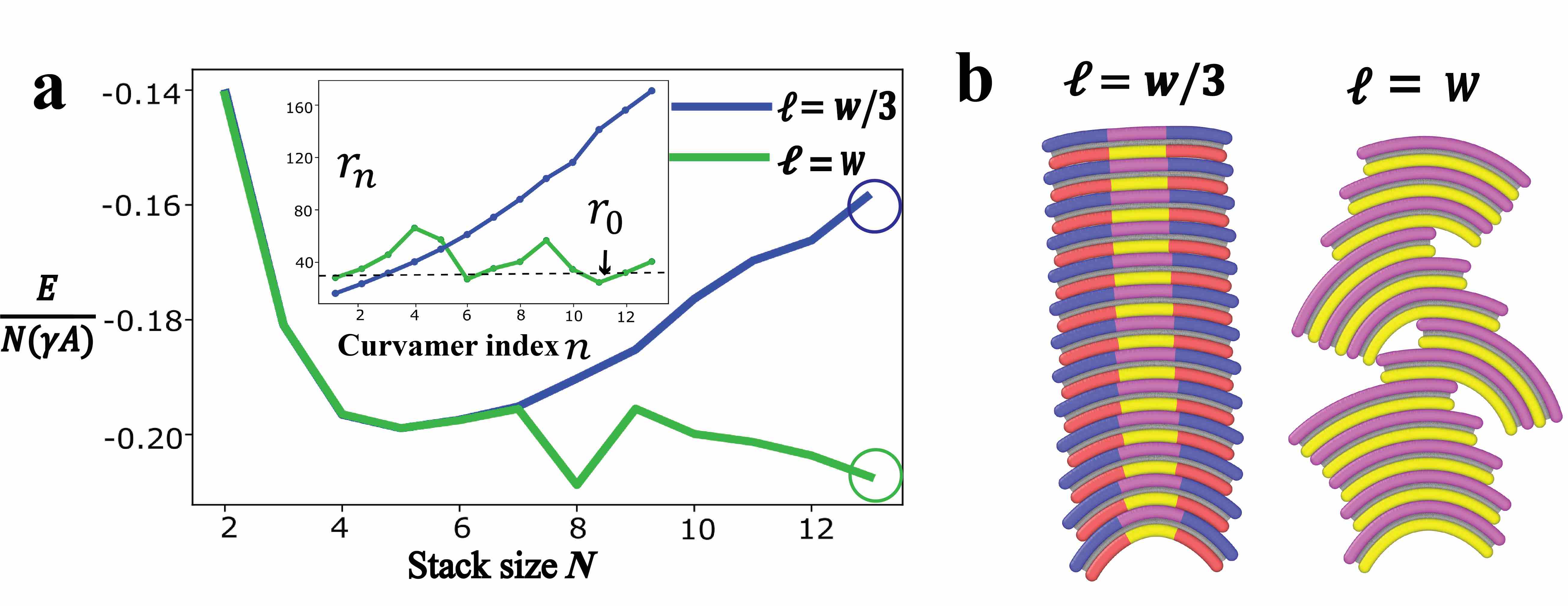}
    \caption{\textbf{(a)} The energy density plots for two interaction geometries with different patch lengths for curvamers with $\sigma = 0.06t$, $r_0=3.5t$ and $S =0.07$. For $\ell=w/3$ (blue) a minimum indicates that assembly is self-limited with an optimal stack size $N_{\rm min}=5$, while for $\ell=w$  (green) energy density decreases below a metastable minimum at large $N$. The radii of curvature in the corresponding stacks of $N=13$ are plotted in the inset, showing the contrast between monotonic curvature focussing through the $\ell=w/3$ stack, and the oscillating curvature profile of of the $\ell=w$ stack. \textbf{(b)} Comparison on $N=13$ stack configurations for $l=w/3$ and $\ell=w$.}
    \label{fig:escape2}
 \end{figure}
 
 The second mode of escape develops by ``misalignment" of curvamer binding, and is observed when the size of binding patch $\ell$ is increased, permitting low-cost lateral sliding of bound curvamers.  In Figure~\ref{fig:escape2} we compare assembly for short-range adhesion ($\sigma/\delta=0.22$) for particles with a narrow ($\ell = w/3$) and broad ($\ell = w$) adhesive binding patch.  In particular, we consider the energetic ground states resolved when subjecting the initially aligned assemblies to simulated annealing at finite temperature (see SI Sec. 5).  Notably, the curvamer stacks with narrow binding patches retain their alignment, such that curvature focussing propagates throughout the stack height.  In comparison, assemblies of curvamers with broad binding patches become unstable to lateral rearrangement between adjacent curvamers in the stack.  This results in large complex ``super -stacks'' composed of looser assembly of multiple aligned and concentric ``substacks'' of $\sim 2-4$ units.  The lateral sliding between adjacent sub-stacks effectively redirects the curvature focussing to outside of the assembly, allowing the super-stack to grow larger without generating superextensive elastic costs for shape change.  
  
In summary, we have demonstrated that geometry of curvature focussing can be used to design frustration-limited assemblies of curved colloidal particles, and crucially, explored how particle-scale features controlling interactions inhibit or allow the assembly to escape the thermodynamic consequences of geometric frustration. Taken together, these results point to critical features required for achieving large self-limiting dimensions:  (i) inter-particle adhesion that is effectively ``stiffer'' than required intra-particle deformation and (ii) interactions that maintain alignment of curvature frustration throughout the assembly.  While exploiting frustration to realize size-controlled assembly is still an emerging concept, it is clear that self-limitation of curvamer assemblies offers important advantages.  Namely, the size of the self-limiting curvamer stack can reach especially large values in comparison to the single particle size.  For example, a recent experimental design of incommensurate DNA origami particles (`PolyBricks') reports self-limiting chains of mean length $\le 5$ particles or less \cite{berengut_self-limiting_2020}, while simulations of frustrated tubules reported free energy minima only up to $\sim 4-8$ particle lengths in dimension \cite{tyukodi_thermodynamic_2021}.  This limitation on the ``escape size'' of assembly derives from the generic competition between elastic costs of accumulating frustration versus ``flattening out'' misfit in an infinite assembly.  In curvamer assembly, the latter cost exceeds the former until the stack thickness reaches $\sim r_0$, which can be made arbitrarily larger simply by decreasing precurvature and accounts for the large range of self-limiting equilibria ($N_{\rm min} \sim 3-70$) exhibited in our model. 

We conclude by briefly noting that this large range of self-limiting sizes is expected to be accessible with experimental colloidal systems.  Considering an adhesive attraction of $\gamma A \approx 10 ~k_B T$  between shell-like colloids with $w \sim 5 ~ {\rm \mu m}$ (achievable by e.g. short-range depletion interactions \cite{mason_osmotically_2002}) and $t \sim 50 ~ {\rm nm}$ for elastomeric curved shells, we expect a bending stiffness in the range $B \sim 10^3 k_B T$ \cite{bae_measuring_2015}. For pre-curvature values $\kappa_0 t \sim 10^{-3}-10^{-1}$, our model predicts self-limitation up to hundreds of times larger than the subunit size itself.

\begin{acknowledgments}
We are grateful to K. Sullivan and C. Santangelo for extensive discussions on this work. This work was supported by the Brandeis Center for Bioinspired Soft Materials, an NSF MRSEC, DMR-2011846 (NT, DMH, RCH, GMG), and through NSF grant No. DMR-2028885 (DMH, GMG).
 \end{acknowledgments}

\bibliographystyle{apsrev4-1}
\bibliography{curvamer}

\end{document}